\documentclass[11pt,a4paper]{article}
\pdfoutput=1

\usepackage{pgf}
\usepackage{tikz}
\usepackage{graphicx}
\usetikzlibrary{arrows,automata}
\usepackage{amssymb,amsmath,amsfonts}
\usepackage{longtable}
\usepackage{verbatim}
\usepackage{color}
\usepackage[mathscr]{euscript}
\usepackage{framed}
\usepackage{mdframed}
\usepackage{amsmath}
\usepackage{amsfonts}
\usepackage{mathrsfs}
\usepackage{amssymb}
\usepackage{mathrsfs}  
\usepackage{subcaption}
\usepackage{caption}

\usepackage[left=26mm,top=20mm,right=26mm,bottom=20mm]{geometry}

\usepackage{color}
\usepackage{cite}
\usepackage{hyperref}
\hypersetup{colorlinks, citecolor=bluscuro, linkcolor=black, urlcolor=bluscuro}
\definecolor{rossos}{cmyk}{0,1,1,0.55}
\definecolor{bluscuro}{rgb}{0.15, 0.2, .85}
\definecolor{bluchiaro}{cmyk}{1,.3,0.,0.1}

\graphicspath{{./Figures/}}

\newcommand{\be}{\begin{equation}}
\newcommand{\ee}{\end{equation}}
\newcommand{\bea}{\begin{eqnarray}}
\newcommand{\eea}{\end{eqnarray}}
\newcommand{\beq}{\begin{equation}}
\newcommand{\eeq}{\end{equation}}

\def\beqa{\begin{eqnarray}}

\def\eeqa{\end{eqnarray}}

\def\lsim{\mathrel{\rlap{\lower4pt\hbox{\hskip0.5pt$\sim$}}
    \raise1pt\hbox{$<$}}}         
\def\gsim{\mathrel{\rlap{\lower4pt\hbox{\hskip0.5pt$\sim$}}
    \raise1pt\hbox{$>$}}}         

\usepackage[normalem]{ulem}

\newcommand{\arXiv}[2]{\href{http://arxiv.org/pdf/#1}{{\tt [#2/#1]}}}

\begin{document}

\vspace{0.1in}

\begin{center}
{\Large\bf\color{black}
Gravitational Waves from Density Perturbations in an Early Matter Domination Era
}\\
\bigskip\color{black}
\vspace{.5cm}
{ {\large Ioannis Dalianis ${}^{a b}$ and  Chris Kouvaris $^{a}$}
\vspace{0.3cm}
} \\[5mm]
{\it {$^a$\, Physics Division, National Technical University of Athens, \\ 15780 Zografou Campus, Athens, Greece \\
$^b$ Department of Physics, University of Athens, \\ University Campus, Zographou 15784, Greece}} \, \\ \vspace{0.2cm}
E-mail : dalianis@mail.ntua.gr,\, kouvaris@mail.ntua.gr
\\[2mm]

\end{center}

\vskip.2in

\noindent
\rule{16.6cm}{0.4pt}

\vspace{.3cm}
\noindent
{\bf \large {Abstract}}
\vskip.15in 
We calculate the gravitational wave background produced from density perturbations in an early matter domination era where primordial black holes  form. The formation of black holes requires perturbations out of the linear regime. Space with such perturbations reach a maximum expansion before it collapses asymmetrically forming a Zel'dovich pancake which depending on the parameters can either lead to  a black hole or a virialized halo. In both cases and due to the asymmetry of the collapsing matter, a quadrupole moment generates gravitational waves which  leave an imprint in the form of a stochastic background that can be detectable by near future gravitational interferometers.

\noindent


\noindent

\bigskip

\noindent
\rule{16.6cm}{0.4pt}

\vskip.4in

\section{Introduction}

With the advent of gravitational wave (GW) detection from mergers of black holes \cite{Abbott:2016blz}, a new era of exploration has begun that could accelerate our progress in understanding black hole physics and the conditions of the early universe. Apart from pure astrophysical processes like the merger of two black holes or neutron stars, gravitational interferometers could also probe the early universe for example by observing mergers of primordial black holes or by observing  a stochastic gravitational wave background produced from process much earlier than the formation of the first stars. The LIGO collaboration has already produced upper limits in such stochastic backgrounds~\cite{TheLIGOScientific:2016dpb}. This will be further extended in the near and not so near future, by a network of operating and designed gravitational wave detectors that will probe a vast range of different frequency bands.  Pulsar time array (PTA) GW experiments \cite{Chen:2019xse} have a sensitivity at the nHz frequency band, space-based designed interferometers like LISA \cite{Audley:2017drz}, Taiji \cite{Guo:2018npi}, Tianqin \cite{Luo:2015ght}, Decigo \cite{Seto:2001qf, Sato:2017dkf} are mostly sensitive to milli- and deci-Hz frequency bands and the LIGO/Virgo and  Einstein telescope \cite{Sathyaprakash:2012jk} ground based interferometers are sensitive to larger frequencies.
 
Several physical sources that generate GWs in the early universe have been proposed  so far~\cite{Sasaki:2018dmp, Kuroyanagi:2018csn, Christensen:2018iqi, Caprini:2018mtu, Sa:2011rm}.  Among them there is an unambiguous source: the primordial density scalar perturbations that source tensor modes at non-linear order perturbation theory \cite{Matarrese:1992rp, Matarrese:1993zf, Matarrese:1997ay, Noh:2004bc, Carbone:2004iv, Nakamura:2004rm}, the so-called induced or secondary gravitational waves.
Density perturbations in the non-linear regime induce tensors with amplitudes proportional to the square of the density perturbations.  
The induced tensors are suppressed 
by the small  $\delta\rho/\rho $ value at the CMB scales \cite{Aghanim:2018eyx},
 but might be sizable if the primordial curvature perturbations are enhanced at small scales.  
 A potential  detection of the relic GW stochastic background might be a direct probe of the very early  cosmic history, which is unknown for $t\lesssim 1$s \cite{Allahverdi:2020bys}.

The features of the induced GWs are  well understood and described if the associated gravitational potential 
decays with time. This is the case for the radiation domination era \cite{Mollerach:2003nq, Ananda:2006af, Baumann:2007zm, DeLuca:2019llr};  results also  exist  for a background with non-thermal, but non-zero, pressure \cite{Domenech:2019quo, Domenech:2020kqm, Dalianis:2020cla}.
On the other hand, the gravitational potential during a matter dominated universe remains constant and the density perturbations grow.
Ref. \cite{Assadullahi:2009nf} pointed out the limitations of the perturbative analysis due to the presence of a non-linear scale. Inside the perturbative regime, Ref. \cite{Inomata:2019ivs, Inomata:2019zqy}   
performed a thorough  investigation where the role and contribution of the subhorizon perturbations has been stressed.
However, the non-linear scale  restricts the applicability of the existing formalism only on a short period before the reheating of the universe and the transition into the radiation era. 
In addition, 
the induced tensors are generated from quadratic combinations of first order scalars and during matter era are continuously sourced and thus do not propagate freely; this fact raises an issue of gauge dependence, discussed in \cite{Hwang:2017oxa,Tomikawa:2019tvi,  Gong:2019mui, DeLuca:2019ufz, Inomata:2019yww, Chang:2020iji,Chang:2020mky, Domenech:2020xin}.

In this work, motivated by the above theoretical and analytical limitations  we choose  to tackle the problem of the computation of the GW signal during an early matter domination era (eMD) following 
an alternative analytic yet nonlinear approach. Studying nonlinear gravitation instability is notoriously difficult even within the Newtonian context. 
 However, the pioneering work of Zel'dovich~\cite{Zeldovich:1969sb} have shown that there can be an approximation valid well inside the nonlinear regime (although eventually breaks down too) \cite{Bernardeau:2001qr}. The Zel'dovich approximation describes the nonlinear evolution of pressureless density perturbations which deviate from the sphere. As the system evolves, an initial deviation from the spherical shape becomes more pronounced as times goes on, leading to flattened structures the so-called Zel'dovich pancakes. This increasingly asymmetric evolution of the collapsing perturbation creates substantial quadrupole moment which in turn creates GWs. The approximation breaks at some point and the eventual fate of the collapsing perturbation is dictated by the final degree of asymmetry of the pancake. If the final configuration (i.e. the pancake) satisfies the hoop conjecture~\cite{Thorne:1972ji, Misner:1974qy}, a black hole is formed out of this collapsing  chunk of matter. If the hoop conjecture is not satisfied, the pancake will go through a phase of shell crossing and oscillation due to its asymmetry redistributing the energy among the different particles leading via violent relaxation to a state of a virialized halo~\cite{LyndenBell:1966bi}. In this paper we focus on the first part of the process, i.e. the nonlinear collapse of the perturbation within the Zel'dovich approximation and we will not address here the background that could potentially exist due to the violent relaxation of the Zel'dovich pancakes.
Since our approach is based on the aforementioned nonlinear treatment of the perturbation, we circumvent  the  limitations of the perturbation theory which is strictly valid in the linear regime.  

We note that the GW emission during an eMD era and in the non-linear regime has been discussed in the past in Ref. \cite{Jedamzik:2010hq},
 and later in Ref.\cite{Nakama:2020kdc}.  The early work \cite{Jedamzik:2010hq} gave a description of the non-linear phases until the moment of reheating and an order of magnitude estimation for the total GW signal. Here, aiming at a refined result,  we derive a new  expression for the  spectrum of the GW signal  produced during the collapsing phase  following the Zel'dovich approximation.

Our method provides semi-analytic results that complements the GR perturbative approach in the non-linear regime and  lightens the underlying physical processes 
 helping thus to 
identify the physical quantities and pin down the signatures of the GWs. 
  In addition to the GW spectra we examine the abundance and distribution of the associated primordial  black holes (PBHs).
We examine scenarios that predict a sizable amount of both PBHs and GW background,  and scenarios that predict a significant amount of GWs with the PBH counterpart being either negligibly small or promptly evaporating and we manifest the connection between the PBH mass and the GW counterpart. Our results can be used to test already existing and viable inflationary models in which PBHs are produced in an eMD phase  \cite{Dalianis:2018frf, Dalianis:2019asr, Ballesteros:2019hus}.

It is worth stating that an eMD phase of the universe is well motivated in several beyond Standard Model scenarios. Such a phase happens before the universe is dominated by radiation and it can practically last almost up $t\simeq1$ sec which is the start of Big Bang Nucleosynthesis (BBN).

The paper is organized as follows. In section II we describe the evolution of a  density perturbation characterized by a small deviation from sphericity.
In section III we derive the general formula for the spectral energy density  of the GWs. 
In section IV we estimate the GW signal and amplitude following analytic steps and approximations. In section V we connect the GW signal with the associated PBH abundance. We conclude in section VI.

\section{Evolution and collapse of the overdensities in eMD} \label{SecMethod}

As we mentioned in the introduction we will not use second order perturbation theory in order to find the GW induced by the overdensities, but we will estimate the quadrupole moment of pressureless collapsing matter within the Zel'dovich approximation.
Here we will closely follow the detailed analysis of \cite{Harada:2016mhb}.  
Within the Zel'dovich approximation, the coordinate of a particle is written as
\begin{equation}
r_i=a(t)q_i +b(t)p_i(q_i),
\label{coord}
\end{equation}
where $a(t)$ is the usual scale factor encoding the expansion of the Universe, $q_i$ is the comoving coordinate,  $b(t)$ is a  growing mode encapsulating the gravitational instability in a pressureless eMD universe and $p_i$ are deviation vectors that depend on the initial perturbation. In order to study the motion of a group of particles around $q_i$, a deformation (strain) tensor $D_{ik}$ is needed defined as  
\begin{align}
D_{ik} & =\frac{\partial r_i}{\partial q_k}=a(t)\delta_{ik}+b(t)\frac{\partial p_i}{\partial q_k} \\
& =\text{diag}(a-\alpha b, a-\beta b, a-\gamma b)\,,
\end{align}
where we have chosen our comoving coordinate system so that the matrix $\partial p_i/\partial q_k$ is diagonal, 
\begin{equation}
\frac{\partial p_i}{\partial q_k}=-\text{diag}(\alpha, \beta, \gamma) \,.
\end{equation} 
A perturbation with  size $q$  and wavenumber  $k=q^{-1}$ enters the horizon at $t_q$ satisfying

\begin{equation}
a(t_q)q=H^{-1}(t_q).
\label{horizon}
\end{equation}
We assume initially small deviations from sphericity, $\alpha b(t_q)/a(t_q) \ll1$, $\beta b(t_q)/a(t_q) \ll1$, $\gamma b(t_q)/a(t_q) \ll1$ so that we can say that at $t_q$ nearly the whole ellipsoid is inside the Hubble sphere, $r_i(t_q)\approx  a(t_q)  q$, where $\vec{r}$ is the 
position of the ellipsoid boundary.
The  mass contained in the overdensity is conserved i.e.,
\begin{equation}
M=\int \rho a^3 d^3r =\bar{\rho}a^3 \int d^3q\,,
\label{mass}
\end{equation}
leading to the following relation between densities
\begin{align}
\rho (a-\alpha b)( a-\beta b)(a-\gamma b)=\bar{\rho}a^3.
\label{Vellp}
\end{align}
The energy contrast after truncating terms beyond the linear order reads
\begin{equation}\label{dlin}
\delta_L \equiv \left(\frac{\rho-\bar{\rho}}{\bar{\rho}}\right)_L=(\alpha+\beta+\gamma) \frac{b}{a}
\end{equation}
where 
$b(t)\alpha \ll a(t)$, $b(t)\beta \ll a(t)$, $b(t)\gamma \ll a(t)$. 
At the moment of horizon entry, the density contrast field is  
\begin{equation}
 \delta_L (t_q) =(\alpha+\beta+\gamma) \frac{b(t_q)}{a(t_q)} \ll 1 ~.
 \label{ratio1}
\end{equation}
Shrinking takes place  for positive $\delta_L>0$.
When $\delta_L \rightarrow 1$ the expression (\ref{dlin}) stops  being valid.
It is known \cite{Mukhanov:2005sc} that during matter domination the density perturbation in the linear regime grows as the scale factor, $\delta \propto a$. Hence,
\begin{equation}
b \propto a^2 \,,
\end{equation}
signifying that  the growing mode $b(t)$ of the deformation tensor describing the collapse, changes much more rapidly than the Hubble flow factor $a(t)$.
The perturbation will collapse at least along one of the three axes and here we assume that 
\begin{equation} \label{hiera}
\alpha>0\, , \quad -\infty<\gamma \leq \beta \leq \alpha<\infty  \quad \text{and}\quad \alpha+\beta+\gamma >0.
\end{equation}
From Eqs.~(\ref{horizon}) and (\ref{mass})  one can relate the horizon entry time $t_q$ with the mass contained
 \begin{equation}
 t_q=\frac{4GM}{3}.
 \end{equation}
 We define four important moments:
The horizon entry $t_q$, the maximum expansion time $t_\text{max}$,  the collapse time $t_\text{col}$ and the (potential) BH formation time $t_\text{BH}$. 
At the moment of maximum expansion, that we labeled  $t_\text{max}$, the mass is about to shrink. By definition it is $\dot{r}_1(t_\text{max})=0$ and one finds $\dot{a}(t_\text{max})=\alpha \dot{b}(t_\text{max})$. 
This implies $\dot{b}/\dot{a}=1/\alpha$ at the maximum expansion
and
\begin{equation}
r_1(t_\text{max})=a(t_\text{max})q_1-\alpha b(t_\text{max})q_1 =  \frac12 a(t_\text{max})q_1,
\end{equation}
which means that at maximum expansion the 1st-axis, $r_1$,  has  half of the size the primordial perturbation would have due to the Hubble flow only.
Since $b\propto a^2$, $\dot{b} \propto 2a\dot{a}$ one finds
\begin{equation} 
\frac{b(t_\text{max})}{a(t_\text{max})}=\frac{1}{2 \alpha}\,.
\label{ratio2}
\end{equation}
Since we know the ratio of $b(t)/a(t)$ at both $t_q$ and $t_{\text{max}}$ from Eqs.~(\ref{ratio1}) and (\ref{ratio2}), we can write
\begin{equation}
\frac{a(t_{\text{max}})}{a(t_q)}=\frac{\alpha+\beta+\gamma}{2 \alpha  \delta_L(t_q)}.
\end{equation}
Taking into account  that $a(t) \propto t^{2/3}$ during matter domination leads us to
\begin{equation} \label{tmax}
t_\text{max}=\left( \frac{\alpha+\beta+\gamma}{2\alpha \delta_L(t_q)} \right)^{3/2} t_q \,.      
\end{equation}

\subsection{Pancake collapse of the overdensity}
Collapse takes place at the time $t_\text{col}$ where $r_1(t_\text{col})=0$.
It is $a(t_\text{col})q_1=\alpha b(t_\text{col})q_1$ and the ratio of the scale factor and the linearly growing mode $b(t)$, 
\begin{equation}  \label{Rcol}
\frac{b(t_\text{col})}{a(t_\text{col})}=\frac{1}{\alpha}\,.
\end{equation}
This is a pancake collapse because the mass becomes a two-dimensional ellipse with the semi-minor and semi-major axes given by 
$r_2(t_\text{col})$ and $ r_3(t_\text{col})$.
The relation $b\propto a^2$ implies
 then
\begin{equation}
\frac{b(t_\text{col})}{b(t_\text{max})} =\left(\frac{a(t_\text{col})}{a(t_\text{max})}\right)^2,
\end{equation}
and from Eq. (\ref{Rcol}) one finds  $a(t_\text{col})=2a(t_\text{max})$.
The scale factor changes with the cosmic time as  $a\propto t^{2/3}$, hence 
\begin{equation}
t_\text{col}=2^{3/2} t_\text{max} \simeq 2.8 \, t_\text{max}.
\end{equation}
The semi-minor and semi-major axes of the overdensity  at the time of collapse depend essentially on the values of the deformation parameters $\alpha, \beta$ and $\gamma$,
\begin{align} \label{panc1}
 r_2(t_\text{col})& =4 \, r_1(t_\text{max})\left( 1-\frac{\beta}{\alpha} \right)= 2 \, a(t_\text{max}) q \,  \left( 1-\frac{\beta}{\alpha} \right),
\end{align}
\begin{align} \label{panc2}
 r_3(t_\text{col}) & =4 \, r_1(t_\text{max})\left( 1-\frac{\gamma}{\alpha} \right)=2 \, a(t_\text{max})q  \, \left( 1-\frac{\gamma}{\alpha} \right).
\end{align}
We see that for $\beta\sim0$ and $\gamma\sim 0$ the non-collapsed other two dimensions might have grown nearly freely due to the  background expansion having size  $a(t_\text{col}) q$.
Also for $\beta$ and $\gamma$  not much smaller than the $\alpha$, i.e. $\alpha\sim \beta\sim \gamma$ we have a spherical collapse  and a black hole might form. 

 Let us mention that, except for the time of maximum expansion for the $r_1(t)$ axis, one can find the time of maximum volume. 
For the special case $\beta\sim \gamma \sim 0$  the equation $\dot{V}_\text{ellip}(t_\text{maxV})=0$ implies that
 $b/a=3/(4\alpha)$ and one finds $t_\text{maxV}\sim (3/2)^{3/2} t_\text{max}$. 
The moment $t_\text{maxV}$ is the moment of minimum density value of the perturbation and signifies a turnover.  It is   $\dot{\rho}_\text{ellip}(t_\text{maxV})=0$ and afterward the density increases. We see that at least until that time the Zel'dovich approximation is reliable and any error at the estimation of $t_\text{col}$ should be of order ${\cal O}(1)$.

The $t_\text{col}$ can be calculated for particular values of the  parameters $\delta_L(t_q)$, $t_q$ and $\alpha, \beta, \gamma$, i.e $t_\text{col}=t_\text{col}(\alpha, \beta, \gamma, \delta_L, t_q)$.
The $\delta_L(t_q)$ is specified by the amplitude of the power spectrum peak; $t_q$ is specified by the position of the curvature  power spectrum peak; $\alpha, \beta, \gamma$ are specified by the Doroshkevich probability function that we introduce right below.

\subsection{Probability distribution of  parameters of the deformation tensor}

\begin{figure}[!htbp]
  \begin{subfigure}{.5\textwidth}
  \centering
  \includegraphics[width=.95 \linewidth]{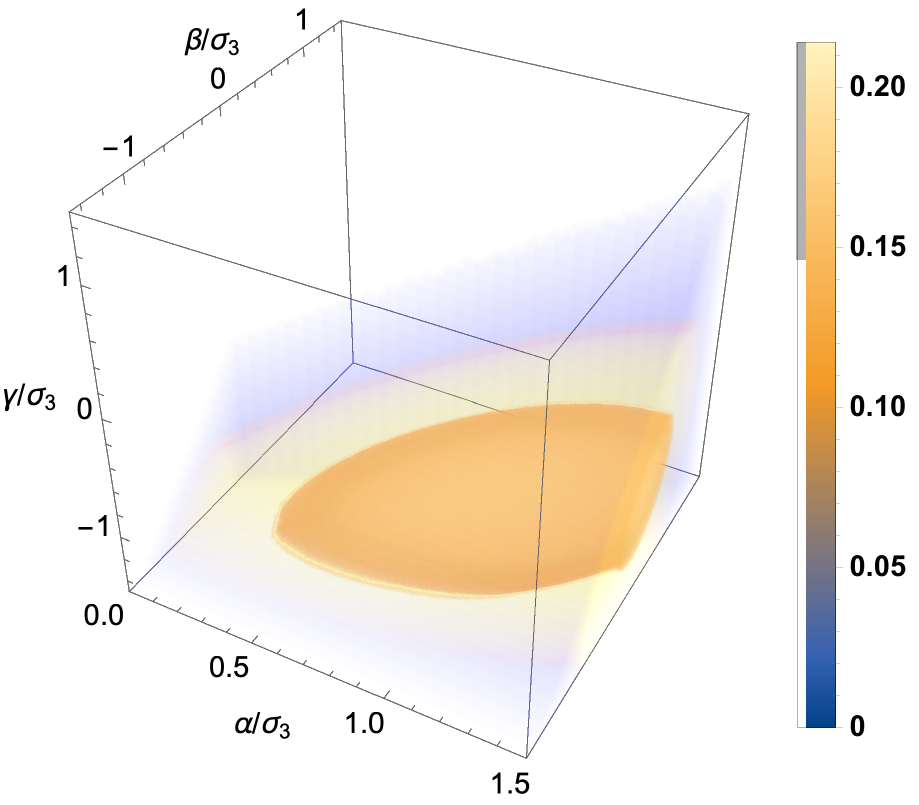}
\end{subfigure}  
  \begin{subfigure}{.5\textwidth}
  \centering
  \includegraphics[width=.95\linewidth]{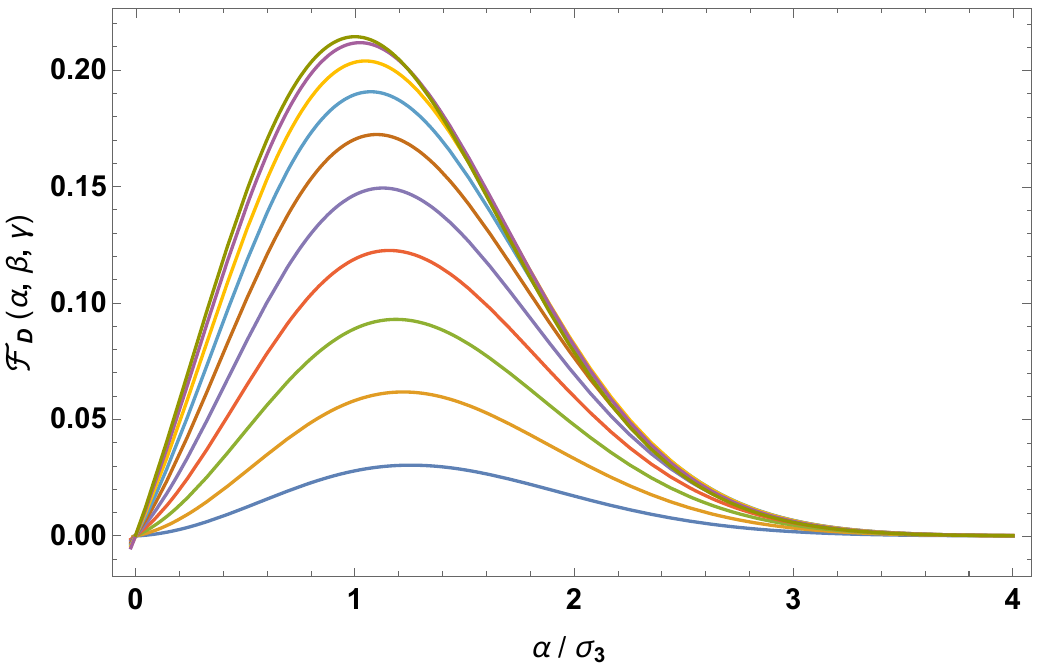}
\end{subfigure}
 \caption{ \label{fig:Dor}~~  
{\it Left  panel:} Density plot for the Doroshkevich probability density ${\cal F}_\text{D}$ for $\alpha, \beta, \gamma$ normalized by the variance $\sigma_3$.
{\it Right Panel:} One dimensional section of the Doroshkevich probability density with respect to the parameter $\alpha/\sigma_3$  for ten different  $\gamma/\sigma_3$ values from $\gamma=-0.1\sigma_3$ (lower curve) to $\gamma=-\sigma_3$ (upper curve) and $\beta/\sigma_3=0$. 
  }
\end{figure}

The Doroshkevich probability distribution of $\alpha$, $\beta$ and $\gamma$ encapsulates and quantifies the finite probability that the considered overdensity deviates from sphericity. It reads~\cite{Doroskevich1970},
\begin{align} \label{DoroProb}
{\cal F}_\text{D}(\alpha, \beta, \gamma, \sigma_3) d\alpha d\beta d\gamma =& -\frac{27}{8\sqrt{5}\pi \sigma^6_3}\times \nonumber 
 \exp\left[- \frac{3}{5 \sigma_3^2}\left((\alpha^2+\beta^2 +\gamma^2)-\frac{1}{2}(\alpha\beta+\beta\gamma+\gamma\alpha) \right)\right] \nonumber \\
& \times (\alpha-\beta)(\beta-\gamma)(\gamma-\alpha) d\alpha d\beta d\gamma,
\end{align}
for $\alpha\geq \beta\geq \gamma$ and is normalized to one, 
\begin{align}
\int_{-\infty}^\infty  d\alpha \int_{-\infty}^\alpha  d\beta  \int_{-\infty}^\beta  d \gamma\,{\cal F}_\text{D}(\alpha, \beta, \gamma, \sigma_3)  =1
\end{align}
The $\sigma$ is the standard deviation of $\delta_L(t_q)$, 
given by the expectation value of the density perturbation at the moment of entry
\begin{equation}
\sigma^2=\left\langle \delta^2_L(t_q)\right\rangle=\left\langle(\alpha+\beta+\gamma)^2\right\rangle  \left( \frac{b}{a}\right)^2(t_q)=5\sigma^2_3,
\end{equation}
after fixing the normalization of $\beta$ as $b(t)/a(t)=a(t)/a(t_i)$.
The probability that the sum of the difference between $\alpha, \beta$ and $\gamma$ takes a value much larger than $\sigma_3=\sigma/\sqrt{5}$ is exponentially suppressed. The most probable values are
 \begin{align} \label{vev1}
     \alpha={\cal O}(\sigma_3), ~ \beta={\cal O}(\sigma_3), ~\gamma={\cal O}(\sigma_3)\,,
 \end{align}
as one can see from Fig. \ref{fig:Dor}; thus  there is a $\sigma_3$ deviation from sphericity. Note that the exact spherical configuration $\alpha =\beta = \gamma$ has zero Doroshkevich probability density.  

In our case  we  add to the hierarchy (\ref{hiera}) for the $\alpha$, $\beta$ and $\gamma$ values the extra constraint that the time of maximum expansion $t_\text{max}$ is larger than the time of entry $t_q$,
\begin{equation}
\alpha+\beta+\gamma >2\, \alpha \,\sigma
\end{equation}
Therefore we constrain the integration region of the  Doroshkevich probability density  in the subspace ${\cal S}$,
\begin{align}  \label{hieraS}
& \, \,\,\quad\quad\quad\quad\quad\quad 0<\alpha<\infty \nonumber \\
&    \, \,\,\quad-\frac{\alpha}{2}(1-2\sigma)< \beta<\alpha    \\
& -\beta-\alpha(1-2 \sigma)< \gamma <\beta  \nonumber
\end{align}
Fig. 1 shows the behavior of the distribution in different  cases, given the hierarchy (\ref{hieraS}). We mention that the triple  integral (\ref{DoroProb}) can be calculated analytically over the $\gamma$ direction and a two dimensional distribution is obtained.

The standard deviation of the density perturbation theory is related to the power spectrum of the density contrast
   $\sigma = \sqrt{{\cal P}_\delta}$.
The density perturbation in the linear regime in the Newtonian gravity coincides with that in the relativistic perturbation theory in comoving slicing both at subhorizon and superhorizon scales in the matter dominated phase \cite{Peebles1980, Hwang:2012bi, Harada:2016mhb}. 
At horizon crossing the variance $\sigma$ is related as  $ \sqrt{{\cal P}_\delta}=({2}/{5})\sqrt{{\cal P_R}}$ where ${\cal P_R}$ is the commonly used, e.g. in inflation model building,  power spectrum of the curvature perturbation defined on comoving hypersurfaces.

\section{Derivation of the energy density of GWs}

The GW emission is described by a multipole expansion of the perturbation $h_{\mu\nu}$ to a background spacetime $g_{\mu\nu}$. 
To lowest order  $h_{ij}(t, d) \propto \ddot{Q}_{ij}(t-d/c)/d$, where $Q_{ij}$ is the mass quadrupole moment and $d$ the distance from the source.
In terms of the moment of inertia the  quadrupole moment is written as
\begin{equation}
Q_{ij}=-I_{ij}(t)+ \frac13 \delta_{ij} \text{Tr} I(t),
\label{Qij}
\end{equation}
where $I_{ij}$, is the moment of inertia tensor\footnote{It is defined as \begin{equation}\nonumber
I_{ij}=\int d^3x \, \rho ( \vec{x}) \left(\delta_{ij}|\vec{x}|^2- x_i x_j  \right).
\end{equation}}. By choosing  the principal axes frame, the moment of inertia tensor is
\begin{align}
I_{ij}=\frac15 M
\left(
\begin{array}{ccc}
r^2_2+r_3^2 & 0 & 0 \\
0 & r_1^2+r_3^2 & 0 \\
0 & 0 & r^2_1+r_2^2
\end{array}
\right),
\label{Iij}
\end{align}
where  $M$ is the constant mass enclosed by the ellipsoid.

Using Eq.~(\ref{coord}) and the fact that $b(t)\propto  a^2(t)$,  we can write the coordinates as a function of time as
\begin{align}
\nonumber
r_1(t)  =\frac{3}{2}t_q^{1/3} t^{2/3}\left(1- \frac{1}{2}\left(\frac{t}{t_\text{max}}\right)^{2/3}\right),
\end{align}
\begin{align}
r_2(t) =\frac{3}{2}t_q^{1/3} t^{2/3}\left(1- \frac{\beta}{2\alpha}\left(\frac{t}{t_\text{max}}\right)^{2/3}\right),
\label{rs}
\end{align}
\begin{align}
\nonumber
r_3(t) =\frac{3}{2}t_q^{1/3} t^{2/3}\left(1- \frac{\gamma}{2\alpha}\left(\frac{t}{t_\text{max}}\right)^{2/3}\right).
\end{align}
where $t_\text{max}=t_\text{max}(\alpha, \beta, \gamma, \delta_L)$.

\subsection{The energy density of GWs}

Let us estimate here  the 
 amount of gravitational waves produced in the eMD era due to a density perturbation with wavelength $q$ that enters the horizon.
At the time of entry $t_q$, the Hubble  comoving volume is $V_q=(4/3) \pi q^3$.
Within the current comoving volume of our universe  $V_\text{com}(t_0)=(4/3)\pi {\cal H}^{-3}(t_0)$ there are ${\cal H}^{-3}(t_0)/q^{3}$  volumes with size $V_\text{q}$.
At the primordial time $t_q$ the density perturbation in each Hubble patch is characterized by a degree of initial asphericity described by the parameters $(\alpha, \beta, \gamma)$ of the deformation tensor and with a distribution of values given by the ${\cal F}_\text{D}$ probability density. 
Taking these into account we define the differential number of sources with deformation parameters $(\alpha, \alpha+d\alpha), (\beta, \beta+d \beta), (\gamma,\gamma+d\gamma)$, 
\begin{align}
dN_\text{}= \frac{V_\text{com}(t_0)}{\frac{4}{3}\pi q^3} {{\cal F}_\text{D}(\alpha, \beta, \gamma, \sigma)\, d\alpha d\beta d\gamma}.
\end{align}
The power emitted in the form of GWs from each Hubble patch, that encloses a perturbation with a quadrupole tensor $Q_{ij}$, is
\begin{equation}
\frac{dE_e}{dt}=\frac{G}{5c^5} \sum_{ij} \dddot{Q}_{ij}(t)\dddot{Q}_{ji}(t).
\end{equation}
Let's define an initial $t_\text{i}$ and a final $t_\text{f}$ moment of the GW emission. 
We separate the interval $[t_\text{i}, t_\text{f}]$ in $N$ intervals of $\delta t$ size and for each interval we can expand the quadrupole in Fourier series,
\begin{equation}
Q_{ij}(t)=\sum_{n=-\infty}^{\infty} \tilde{Q}^N_{ij}(n\omega_0) e^{-in\omega_0 t},
\end{equation}
where $\omega_0=2\pi/\delta t$ and $t~\epsilon~[t_N, t_N+\delta t]$.  The $\tilde{Q}$ components are
\begin{equation}
\tilde{Q}^N_{ij}(n\omega_0)
=\frac{\omega_0}{2 \pi}\int^{t_N+\delta t}_{t_N} Q_{ij}(t) e^{-in\omega_0 t} dt\,.
\end{equation}
By taking the continuum limit for $\omega$ we transform the last two equations to
\begin{equation}
Q_{ij}(t)=\int \tilde{Q}_{ij}^N e^{i\omega t} d\omega,
\end{equation}
\begin{equation} 
\tilde{Q}^N_{ij}(\omega)=\frac{1}{2\pi} \int^{t_N+\delta t}_{t_N} Q_{ij}(t)e^{-i\omega t} dt.
\label{Qomega}
\end{equation}
The emitted energy $E_e$ in the time bin $[t_{N_i}, t_{N_i}+\delta t]$ is
\begin{align} \nonumber
E_e^N &=\frac{2G}{5 c^5} \int^{t_N+\delta t}_{t_N} \sum_{ij}\sum_{nm}(in\omega)^3(-im\omega)^3 
 \times \tilde{Q}^N_{ij}(n\omega) \tilde{Q}^{*N }_{ij}(m \omega)
 e^{-i(n-m)\omega t} dt \nonumber  \\
& =\frac{2G}{5 c^5} \sum_{ij}\sum_{n}(n\omega)^6 \tilde{Q}^N_{ij}(n\omega) \tilde{Q}^{*N }_{ij}(n \omega) \delta t .
\end{align} 
The above expression has a factor of 2 because we now restrict ourselves in positive $n$ and $m$ since the negative ones contribute equally. In the continuum limit 
\begin{equation}
d E^N_e=\frac{4\pi G}{5 c^5} \omega^6 \sum_{ij} |\tilde{Q}^N_{ij}|^2 d\omega,
\end{equation}
and the differential energy emitted per logarithmic interval 
\begin{align} \nonumber
& \frac{dE^N_e}{d\ln \omega}=\frac{4\pi G}{5c^5} \omega^7 \sum_{ij}|\tilde{Q}^N_{ij}(\omega)|^2,
\end{align}
 within the time bin $[t_N, t_N+\delta t]$.
Hence, we find that the infinitesimal energy density today 
of the differential number of sources with deformation parameters $\{(\alpha, \alpha+d\alpha), (\beta, \beta+d \beta), (\gamma,\gamma+d\gamma) \}$ 
is 
\begin{align} 
dE_\text{GW}(\alpha, \beta, \gamma)=&\sum_N \frac{1}{1+z_N} \frac{4\pi G}{5 c^5} \omega^7 \sum_{ij}|\tilde{Q}_{ij}^N(\omega)|^2
 \frac{V_\text{com}(t_0)}{\frac{4\pi}{3} q^3}
{\cal F}_\text{D}(\alpha, \beta, \gamma, \sigma) d\alpha d\beta d\gamma \,d\ln\omega.
\end{align} 
We considered that the path of the gravitons, moving in a radial direction is obtained by setting $ds^2=0$, and a graviton emitted with an energy $E_e$ is today observed with  $E=E_e/(1+z)$ and frequency $\omega_0=\omega/(1+z)$.
The cosmological redshift is given by $1+z=a(0)/a(t_e)$, where we take $a(0)=1$.

Let us  focus here on the GWs emitted from the Hubble patches that experience a pancake collapse of the density perturbation.  We expect that these Hubble patches give the dominant contribution to the
total GW signal.
Inside the Hubble sphere the asphericity will reach the limiting stage of pancake collapse if the time of reheating is sufficiently late, i.e $t_\text{col}<t_\text{rh}$. 
Taking as initial time the turning point of maximum expansion $t_\text{i}=t_\text{max}$ and as final the moment of collapse $t_\text{f}=t_\text{col}$,
we specify the total number of sources as
\begin{align}\label{Ntot}
N_\text{tot} \simeq \int \frac{V_\text{com}(t_0)}{\frac{4}{3}\pi q^3} \Theta\left(t_\text{rh}-t_\text{col}(\alpha, \beta, \gamma)\right){{\cal F}_\text{D}(\alpha, \beta, \gamma, \sigma)\, d\alpha d\beta d\gamma}.
\end{align}
In the unphysical limit $t_\text{rh}\rightarrow \infty$ the number of sources is equal to the total number of Hubble patches, $ {V_\text{com}(t_0)}/({\frac{4}{3}\pi q^3})$.  
Separating the duration $[t_\text{max}, t_\text{col}]$ in $N$ intervals, the differential energy density per observed logarithmic frequency interval and deformation configuration  is 
\begin{align} \nonumber 
\frac{d\rho_\text{GW}(t_0, f_0)}{d\alpha d\beta d\gamma\, d\ln f_0 } & = \sum_N \frac{1}{1+z_N} \frac{4 \pi G}{5 c^5}  \sum_{ij}|\tilde{Q}_{ij}^N\left(2\pi f_0(1+z_N)\right)|^2 \\
&
\times (2\pi f_0(1+z_N))^7\Theta\left(t_\text{rh}-t_\text{col}(\alpha, \beta, \gamma)\right)
\left(\frac{4\pi}{3}q^3 \right)^{-1} {\cal F}_\text{D} (\alpha, \beta,\gamma, \sigma_3). \label{result_rho}
\end{align} 
 The $f_0$ is the frequency parameter today and should not be confused with a specific frequency value. Cosmologists describe the stochastic GW backgrounds by their energy density normalized by the critical energy density of the universe, $\rho_\text{crit}$, the so called  spectral energy density parameter,
\begin{equation}
\Omega_\text{GW}(t, f)
\equiv \frac{1}{\rho_\text{crit}} \frac{d\rho_\text{GW}}{d \ln f}\,.
\end{equation}
Its value at the present time reads
\begin{align} \nonumber 
\Omega_\text{GW}(t_0,f_0) =& \frac{1}{\rho_\text{crit}(t_0)}
\int\int\int_{\cal S}
\,d\alpha d\beta d\gamma\,  \sum_N \frac{1}{1+z_N} \frac{4 \pi G}{5 c^5}  \sum_{ij}|\tilde{Q}_{ij}^N\left(2\pi f_0(1+z_N)\right)|^2 \\
&
\times (2\pi f_0(1+z_N))^7\Theta\left(t_\text{rh}-t_\text{col}(\alpha, \beta, \gamma, \sigma)\right)
 \left(\frac{4\pi}{3}q^3\right)^{-1} {{\cal F}_\text{D} (\alpha, \beta,\gamma, \sigma)}.   \label{result_Om}
\end{align} 
where the integration takes place over the parameters space ${\cal S}$ (\ref{hieraS}). Eq. (\ref{result_Om}) is the {\it main result of this work}. Apparently, the total energy density parameter today is given by the integral $\Omega_\text{GW}(t_0) = \int  \Omega_\text{GW}(t_0, f_0)\, d\ln f_0$.
In the following we will apply our result  estimating the  contribution to the GW from regions that collapse into PBHs and the entire contribution from all the regions that experience a pancake collapse before the reheating of the universe.

\section{Estimation of the GW signal amplitude and spectrum} 

\label{SecResults}

We can now proceed and calculate the sum of the square of the quadrupole moments. In practice it turned out to be simpler to directly calculate the Fourier transform of $\dddot{Q}_{ij}(t)$.
Using Eqs.~(\ref{Qij}), (\ref{Iij}), (\ref{rs}) and (\ref{Qomega}) for a time bin $[t_1, t_2]$ we get, 
\begin{align} \nonumber
   \omega^6 \sum_{ij}|\tilde{Q}^N_{ij}  (\omega)|^2 = &\frac{1}{54 \pi^2} \frac{t_q^{4/3}t_1^{4/3}}{t_\text{max}^{8/3}}M^2
     \left[1+\left(\frac{\beta}{\alpha}\right)^4+\left(\frac{\gamma}{\alpha}\right)^4
    -\left(\frac{\gamma}{\alpha}\right)^2-\left(\frac{\beta}{\alpha}\right)^2\left(1+\left(\frac{\gamma}{\alpha}\right)^2\right)\right]  \nonumber \\
    & \times \left|\text{Ei}\left[\frac13, i\omega t_1\right]- \left (\frac{t_2}{t_1} \right )^{2/3} \text{Ei}\left[\frac13, i\omega t_2\right] \right|^2 \,.
    \label{FT2}
\end{align}
For the expression of Eq.~(\ref{result_Om})  to be absolutely  correct, $N$ must be large i.e., we should partition the time interval $[t_{\text{max}}, t_\text{col}]$ to a large number of small time intervals. The reason for having to do that instead of taking the Fourier transform for the full interval is because frequencies produced at different times are redshifted by  different amounts as they reach our detectors today.  As a first approximation we will consider only one interval (i.e., $N=1$) and we will consider as if the GWs were emitted instantaneously at $t_\text{col}$. This is a very good approximation  since $t_\text{col}=2^{3/2} t_\text{max}$ and $t_\text{max}$ are relatively close to each other, and therefore the introduced horizontal error in the frequency is at most a factor of 2. By use of 
 Eq. (\ref{FT2}) within this ``average'' approximation,
  Eq. (\ref{result_Om}) is recast into
 \begin{align}\label{result_Om1}
  \Omega^{\text{}}_\text{GW}(t_0,f_0) &= \frac{1}{\rho_\text{crit}}\,   \left({\frac{4\pi}{3}q^3}  \right)^{-1} \frac{4 \pi G}{5c^5}\frac{2\pi f_0}{54\pi^2}
  M^{2} \, \sigma^2
\int\int\int_{\cal S}
 \,d\alpha d\beta d\gamma\, 
  \left(\frac{2\alpha}{\alpha+\beta+\gamma} \right)^2  \times 
  \nonumber
  \\
  &\left[1+\left(\frac{\beta}{\alpha}\right)^4+\left(\frac{\gamma}{\alpha}\right)^4
    -\left(\frac{\gamma}{\alpha}\right)^2-\left(\frac{\beta}{\alpha}\right)^2\left(1+\left(\frac{\gamma}{\alpha}\right)^2\right)\right] \times  \nonumber \\
 &  \left|\text{Ei}\left[\frac13, it_\text{max} \left(\alpha, \beta, \gamma, \sigma)\right) 
  2\pi f_0(1+z_\text{col}\left(\alpha, \beta, \gamma, \sigma)\right) \right] \right. \nonumber \\
&  \left.
  - 2\text{Ei}\left[\frac13, it_\text{col}\left(\alpha, \beta, \gamma, \sigma)\right)  2\pi f_0(1+z_\text{col}\left(\alpha, \beta, \gamma, \sigma)\right)\right] \right|^2  \times  \nonumber \\
& \Theta\left(t_\text{rh}-t_\text{col}(\alpha, \beta, \gamma, \sigma)\right)
 {\cal F}_\text{D} (\alpha, \beta, \gamma, \sigma_3).
\end{align}
{\it  The above equation is the one that we use to derive our numerical results}.

The estimation of the $\Omega_\text{GW}(t_0, f_0)$ requires the integration over the parameter space $(\alpha, \beta, \gamma)$.  
 The step function $\Theta\left(t_\text{rh}-t_\text{col}(\alpha, \beta, \gamma)\right)$ 
 is modulated by $t_\text{col}$ and its value has also a probability distribution.
Also the expectation value of the  expression 
$\left\langle((\alpha+\beta+ \gamma)/\alpha)^{3/2}\right\rangle$ has a very weak sensitivity on  $\sigma$  and hence the expectation value of the key moments of maximum expansion and collapse scale  as   $\left\langle t_\text{max} \right\rangle, \, 
 \left\langle t_\text{col} \right\rangle \propto \sigma^{-3/2}$.  
The step function condition $t_\text{rh}>t_\text{col}(\alpha, \beta, \gamma, \sigma)$
can be translated into an upper bound for the reheating temperature as  
\begin{align}
    T_\text{rh} \lesssim 0.2 
     \left(\frac{\alpha \,\sigma}{\alpha+\beta+\gamma} \right)^{3/4}
      \left(\frac{M_\odot}{M} \right)^{1/2} \left(\frac{g_*}{106.75} \right)^{-1/4} \text{GeV},
    \label{reheat}
\end{align}
where $M_\odot$ is the solar mass. If this upper bound is violated, some regions will not collapse into a pancake before the 
transition to the radiation phase. 
For $t_\text{rh}$ not very close to $\left\langle t_\text{col} \right\rangle$ the $\Theta$ step function does not reduce the parameter space ${\cal S}$ to modify   essentially  the   integral (\ref{result_Om1}).

The redshift  experienced by the  GW emitted at the collapse moment has also a probability distribution and  is given by the expression 
\begin{align}
    1+z_\text{col}=\frac{1+z_\text{rh}}{(6\pi G \rho_\text{rh})^{1/3} t^{2/3}_\text{col}}\,.
\end{align}
Also, the comoving radius of the the perturbation at the time of horizon entry, $q$, corresponds to the wavenumber, 
\begin{align}\label{kMgen}
    q^{-1} (M,  T_\text{rh})&=\left(\frac{3M(1+z_\text{rh})^3}{4\pi \rho_\text{rh}} \right)^{-1/3}, 
    \\
&\simeq 1.2 \, \times 10^{10}  \left( \frac{T_\text{rh}}{10^{10}\text{GeV}}
 \right)^{\frac{1}{3}}  
 \left(\frac{M}{M_\odot} \right)^{-\frac{1}{3}}\nonumber 
 {\text{Mpc}^{-1}}\,, 
 \end{align}
where  $\rho_\text{rh}=\pi^2 g_* T_\text{rh}^4/30$ is the energy density at the time of reheating and the redshift $z_{\text{rh}}$ can be easily found using the conservation of entropy.

  \subsection{GWs associated with the PBH formation}

We will first focus  on the collapsing configurations that lead to formation of PBHs. 
  According to the hoop conjecture ~\cite{Thorne:1972ji, Misner:1974qy} a black hole forms when the circumference in every direction is approximately smaller than $2\pi$ times the Schwarschild radius, $r_s=2 GM/c^2$. The hoop ${\cal C}$ of the  pancake is given by the circumference of an ellipse ${\cal C}= 4r_3 (t_\text{col}) E(e)$ where $E(e)$ is the  complete elliptic integral of the second kind and $e$ the eccentricity of the pancake,
  \begin{equation}
e=\sqrt{1-\left(\frac{r_2(t_\text{col})}{r_3(t_\text{col})}\right)^2}  = \sqrt{1-\left( \frac{\alpha - \beta}{\alpha -\gamma} \right)^2}\,.
\end{equation}
 The hoop conjecture for the ellipsis reads ${\cal C} \lesssim 2\pi r_s$, or
\begin{equation} 
h(\alpha, \beta, \gamma)	\equiv  \frac{\cal C}{2\pi r_s} =\frac{2}{\pi} \frac{\alpha-\gamma}{\alpha^2} E\left( \sqrt{1-\left( \frac{\alpha - \beta}{\alpha -\gamma} \right)^2}\right)\, \lesssim 1\,.
\label{hoop}
\end{equation} 
 The  number of GW sources associated with PBH formation is given by a subspace of $(\alpha, \beta, \gamma)$ parameter space,
\begin{align}
N_\text{PBH}& =\frac{V_\text{com}}{\frac{4}{3}\pi q^3} \int  \Theta(1-h(\alpha, \beta, \gamma)) \Theta(t-t_\text{col}(\alpha, \beta, \gamma, \sigma)){\cal F}_D (\alpha, \beta, \gamma, \sigma) d\alpha d\beta d\gamma,  \nonumber
\end{align}
 that fulfills the hoop criterion. Actually the PBH formation criterion $h<1$ implies
\begin{align} \label{PBHpar}
    \alpha-\gamma\lesssim \frac{\pi}{2} \alpha^2=\frac{\pi}{2} {\cal O}(\sigma^2_3)\,,
\end{align} 
thus  it is  $\alpha-\gamma \sim {\cal O}(\sigma^2_3)$, 
$\alpha-\beta \sim {\cal O}(\sigma^2_3)$ and $\beta-\gamma \sim {\cal O}(\sigma^2_3)$. 
 We define the ratio $\epsilon$,
  \begin{align} \label{apprPBH}
     \frac{\alpha-\gamma}{\alpha}
          \sim\frac{\alpha-\beta}{\alpha}\sim\frac{\sigma^2_3}{\alpha}\equiv \epsilon <1\,.
 \end{align}
Since $\alpha \sim \sigma_3$, for $\sigma_3<<1$, it should be also $\epsilon<<1$. Therefore we can expand  Eq. (\ref{result_Om1}) with respect to the small parameter  $\epsilon=\sigma^2_3/\alpha$ up to quadratic order, approximating the brackets of Eq.  (\ref{result_Om1}) as $4\epsilon^2$, 
and recast the expression for the energy density parameter  in the form, 
\begin{align}\label{PBH_OmFull}
 \Omega^{\text{(PBH)}}_\text{GW}(t_0,f_0) & \simeq \frac{1}{\rho_\text{crit}}  \frac{4 \pi G}{5c^5}\frac{2\pi f_0}{54\pi^2}  M^{2}   \left({\frac{4\pi}{3}q^3}  \right)^{-1} \frac{4}{25}\, \sigma^6\, \times \nonumber \\
 & \int\int\int_{{\cal S}_\text{PBH}}
 \,d\alpha d\beta d\gamma\, 
 \left(\frac{2\alpha}{\alpha+\beta+\gamma} \right)^2 
 \frac{1}{\alpha^2}  \nonumber \\
 &  \left|\text{Ei}\left[\frac13, it_\text{max} \left(\alpha, \beta, \gamma, \sigma)\right) 
  2\pi f_0(1+z_\text{col}\left(\alpha, \beta, \gamma, \sigma)\right) \right] \right. \nonumber \\
&  \left.
  - 2\text{Ei}\left[\frac13, it_\text{col}\left(\alpha, \beta, \gamma, \sigma)\right)  2\pi f_0(1+z_\text{col}\left(\alpha, \beta, \gamma, \sigma)\right)\right] \right|^2  \nonumber \\
& \times \Theta\left(t_\text{rh}-t_\text{col}(\alpha, \beta, \gamma, \sigma)\right)
  {\cal F}_\text{D} (\alpha, \beta, \gamma, \sigma_3),
\end{align}
where the ${{\cal S}_\text{PBH}}$ is selected by the hoop-condition step function $\Theta(1-h)$.
In particular, the constraint $h\lesssim 1$ implies that only a subspace of $-\infty<\alpha<\infty $, $-\infty <\beta<\alpha$ and $-\infty<\gamma<\beta$ leads to PBH formation, that in terms of  new variables $x=(\alpha+\beta+\gamma)/3$, $y=(\alpha-2\beta+\gamma)/4$ and $z=(\alpha-\gamma)/2$ reads $0<x$, $-x^2/4<y<x^2/4$,  $2|y|<z<x^2/2$ in the circular limit ($e=0$) and  $0<x$, $-x^2\pi/8<y<x^2\pi/8$,  $2|y|<z<x^2\pi/4$ in the eccentric limit ($e=1$) \cite{Harada:2016mhb}.
In the left panel of  Fig. \ref{fig:Pbh&Entire} the GW signal is depicted for two $\sigma$ values after computing the integral of Eq. $(\ref{PBH_OmFull})$, involving the Doroshkevich probability distribution of non spherical perturbations, over the subspace ${{\cal S}_\text{PBH}}$  in the eccentric limit and for several hundreds of  frequency values over the frequency band of interest.
For different $\sigma$ values a different integration interval is found; 
note that  as the $\sigma$ decreases $\left\langle 1/ \alpha^2  \right\rangle$ also decreases  due to the shrinking of the parameter space that realize PBH formation.

  In principle we should evaluate the quadrupole induced GW signal not up to $t_\text{col}$ but up to $t_\text{BH}$ the first moment where the hoop conjecture was satisfied. Obviously $t_\text{col}>t_\text{BH}$.  In our approximation we have used as the end time of emission   $t_\text{col}$ and  not $t_\text{BH}$. However we will argue that the two times are very close to each other, and therefore using $t_\text{col}$ instead of $t_\text{BH}$ does not introduce any substantial error. Given that at any time $t$ between $t_\text{max}$ and $t_\text{col}$ the circumference of the ellipsoid is given by ${\cal C}= 4r_3 (t) E(e)$, we can rewrite the condition for the hoop conjecture of Eq.~(\ref{hoop}) for time $t$ as
 \begin{equation}
 \frac{2}{\pi}\xi\frac{\alpha -\xi\gamma}{\alpha^2} E(e) \lesssim 1,
  \end{equation}
where $\xi= (t/t_\text{col})^{2/3}$. For $\xi=1$, one recovers Eq.~(\ref{hoop}). In the circular limit $E(0)=\pi/2$ and therefore the above equation becomes 
\begin{equation}
 \xi\frac{\alpha -\xi\gamma}{\alpha^2}  \lesssim 1.
  \end{equation} 
  This equation must be satisfied in order to form a black hole. Given that $\xi$ takes values within $[0.5,1]$ for $t_\text{BH}<t<t_\text{col}$, $\alpha$ and $\gamma$ are ${\cal O} (\sigma_3)$ while $\alpha-\gamma \sim {\cal O}(\sigma_3^2)$, one realizes from the above equation that it cannot be satisfied unless $\xi$ is fine tuned close to 1 in order to make $\alpha-\xi\gamma \sim {\cal O}(\sigma_3^2)$. If  $\xi$ is not very close to 1,  $\alpha-\xi\gamma \sim {\cal O}(\sigma_3)$  and the above equation cannot be satisfied. A similar argument works also for the eccentric limit leading to the same conclusion $t_\text{BH}\simeq t_\text{col}$.

   \subsection{GWs not necessarily associated with  PBH formation}

Let us consider now the entire GW contribution from the overdensities  that enter the horizon at the time of entry $t_q$, evolve and collapse into pancake shape before the reheating of the universe.  
Here we include  the full parameter space for $\alpha, \beta, \gamma$, Eq. (\ref{hieraS}), contrary to the previous case (\ref{PBHpar}), where we limit ourselves to  configurations that collapse into PBHs. There is no surprise that this case leads to larger GW signals. This is easily understood from the fact that PBH formation requires small deviations from sphericity and therefore  smaller parameter space and smaller quadrupole moment which is essential for GW generation. On the contrary collapsing overdensities which are too asymmetric to form PBHs have larger quadrupole moments and larger parameter space to explore. Now unlike the case of PBH formation the difference between the deformation factors is of the order of the variance value,  $\alpha-\gamma ={\cal O}(\sigma_3)$ (and not ${\cal O}(\sigma_3^2))$.

\begin{figure}[!htbp]
  \begin{subfigure}{.5\textwidth}
  \centering
  \includegraphics[width=.95 \linewidth]{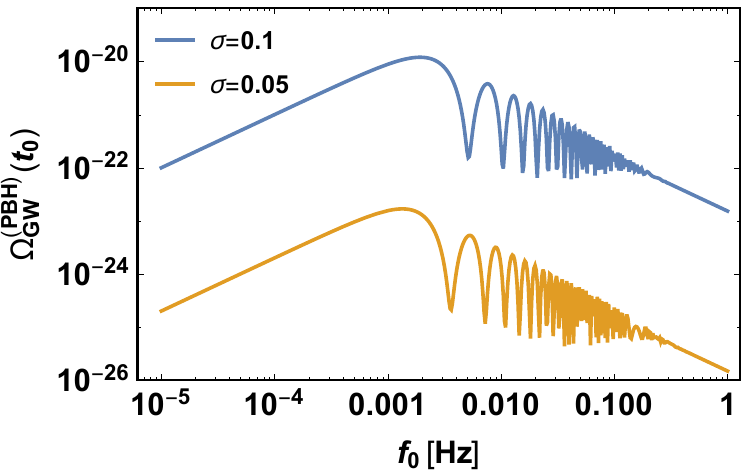}
\end{subfigure}  
  \begin{subfigure}{.5\textwidth}
  \centering
  \includegraphics[width=.95 \linewidth]{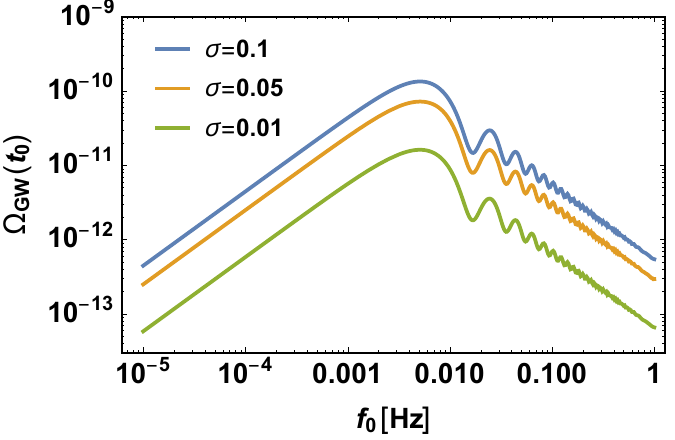}
\end{subfigure}
 \caption{\label{fig:Pbh&Entire}~~  
{\it Left  panel.} The GW signal produced only by the regions that collapse into PBH  for different  $\sigma$ values and reheating temperature $T_\text{rh}= 10^4$ GeV. The horizon mass is $M=10^{-12}M_\odot$.
{\it Right panel.} The entire GW signal for the same horizon mass and temperature.
  }
\end{figure}

\begin{figure}[!htbp]
  \begin{subfigure}{.5\textwidth}
  \centering
  \includegraphics[width=.95 \linewidth]{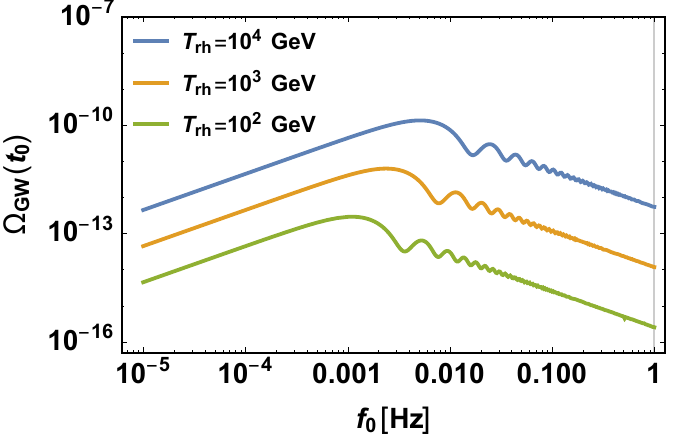}
\end{subfigure}  
  \begin{subfigure}{.5\textwidth}
  \centering
  \includegraphics[width=.95 \linewidth]{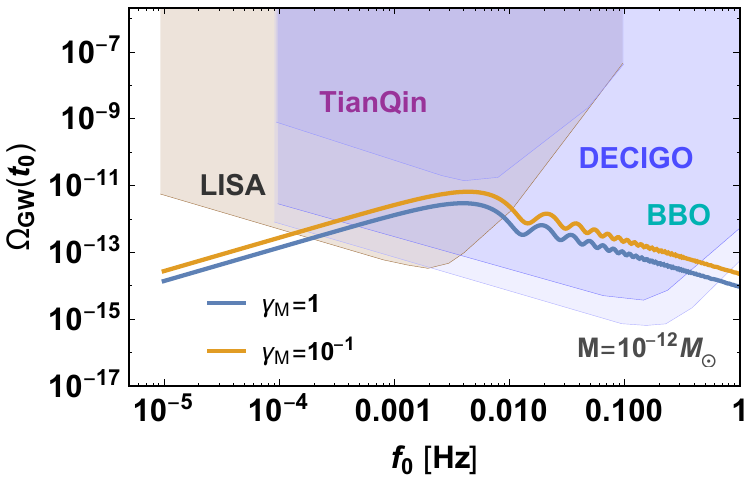}
\end{subfigure}  
 \caption{\label{Fig:TempLISA}~~  
{\it Left  panel.} The GW signal for $\sigma=0.1$ and horizon mass $M=10^{-12} M_\odot$  for three different reheating temperatures.
{\it Right Panel.}   The maximum GW signal in the LISA-BBO frequency band for the PBH dark matter scenario  for two different  $\gamma_\text{M}$ values.
 For $\gamma_\text{M}=1$ it is  $\sigma=0.0047$ 
and for  $\gamma_\text{M}=0.1$ it is $\sigma=0.0071$ so that $\Omega_\text{PBH}=\Omega_\text{DM}$;
 the reheating temperature is chosen to be $T_\text{rh}=T_\text{col}/2$, that is $4.8\times 10^3$ GeV and $6.5\times 10^3$  GeV respectively.
 The horizon and the PBHs have masses $M=10^{-12} M_\odot$ and  $ M_\text{PBH}=\gamma_\text{M} M$ respectively.
  }
\end{figure}

Therefore we find
the spectral energy density parameter of the GWs 
emitted by the Hubble patches that experience a pancake collapse  
after computing the full integral in Eq. (\ref{result_Om1}) over the parameter space ${\cal S}$ given by Eq. (\ref{hieraS}).  We implement the computation after an  integration for several hundreds of  discrete frequency values over a frequency band of interest.   For different values for $M$,  $T_\text{rh}$ and $\sigma$ our numerical results are plotted in Figs. \ref{fig:Pbh&Entire}-\ref{LigoRem}.

The amplitude of the GW signal is  $\sigma^2$ dependent times the integral of the Doroshkevich probability distribution over the ${\cal S}$ parameter space which also depends on $\sigma$. To get a better understanding,
the part of the integrad  in Eq. (\ref{result_Om1}),
\begin{align} \label{apprtotal}
\left[1+\left(\frac{\beta}{\alpha}\right)^4+\left(\frac{\gamma}{\alpha}\right)^4
    -\left(\frac{\gamma}{\alpha}\right)^2-\left(\frac{\beta}{\alpha}\right)^2\left(1+\left(\frac{\gamma}{\alpha}\right)^2\right)\right]   \, \left(\frac{2\alpha}{\alpha+\beta+\gamma} \right)^2 ,
\end{align}
yields a $\sigma$ dependence
roughly inversely proportional to $\sigma$.  Hence the GW signal has a nearly linear dependence on $\sigma$.
By inspecting the frequency dependent terms in Eq. (\ref{result_Om1}) and  after rewriting $2\pi f_0$ as $2 \pi f/(1+z_\text{col})$
 we see that the product  $\omega  \left|\text{Ei}\left[\frac13, it_\text{max} \omega\right]- 2\text{Ei}\left[\frac13, 2\sqrt{2}it_\text{max}  \omega\right] \right|^2 $ features a peak at $\omega \sim 1/ t_\text{max}$  with a value $\sim 1/t_\text{max}$,  where $ \left \langle  t_\text{max} \right\rangle \simeq \,0.17 \, \sigma^{-3/2}\, {t_q}$.
The actual maximum of the GW spectrum is specified  only after computing the  integral in Eq. (\ref{result_Om1})
and  the peak frequency  is given approximately by the relation
\begin{align}\label{fMDpeak} 
 f^\text{}_\text{peak} (M, T_\text{rh})  \simeq   2.5   \times 10^{-8}  
 \left(\frac{M}{M_\odot}  \right)^{-1/3} 
   \left(\frac{T_\text{rh}}{ \text{GeV}} \right)^{1/3}
   \text{Hz}\,.
\end{align}
After normalizing with benchmark values, we write the expression for the spectral energy density parameter  of the GWs at the peak frequency $f_\text{peak}$,
\begin{align}\label{Omegapeak}
\Omega_\text{GW}(f_\text{peak}, t_0)\sim 1.9\times 10^{-7}
\left(\frac{M}{M_\odot}\right)^{2/3} 
\left(\frac{T_\text{rh}}{\text{GeV}}\right)^{4/3}
\left(\frac{\sigma}{0.1}\right)
\end{align}
for $t_\text{rh}>t_\text{col}(\sigma, M)$.

 We stress that the plots in Fig.  \ref{fig:Pbh&Entire}-\ref{LigoRem} are produced after performing the full triple integral in Eq.(\ref{result_Om1}) without assuming any approximation.

The GW spectrum we observe today scales as $\Omega(t_0, f_0)\propto f_0$ in the IR regime and the envelope of the oscillating spectrum falls as $\Omega(t_0, f_0)\propto f_0^{-1}$ in the UV. The peak, which is the peak of the first oscillation, falls much faster, roughly like $f_0^{-3}$, see Fig. \ref{fig:Pbh&Entire}. 
 This is a distinct feature of the GWs produced during eMD from a monochromatic source.
 We  add that  the GW signal produced during an eMD era  (\ref{result_Om}) has also a different dependence on  $\sigma = \sqrt{{\cal P}_\delta}$ and different amplitude 
 compared  to the GW signal  produced during the radiation domination era \cite{ Mollerach:2003nq, Ananda:2006af, Baumann:2007zm} or just before the transition to the radiation era \cite{Inomata:2019ivs, Inomata:2019zqy}.

\section{GWs and the PBH abundance}

In the absence of pressure even minute perturbations will evolve, and in the spherical limit, will collapse into a black hole.   PBH production during matter era  has been first examined in  Ref. \cite{Khlopov:1980mg, Polnarev:1986bi}.
Ref. \cite{Harada:2016mhb} examined the PBH production in a matter dominated universe and considered the non-spherical effects in gravitational collapse.
The production probability of PBH is given by
\begin{equation}
\beta_0=\int_0^\infty d\alpha \int_{-\infty}^\alpha d\beta \int_{-\infty}^\beta \,d\gamma\, \Theta(1-h(\alpha, \beta, \gamma)){\cal F}_\text{D}(\alpha, \beta, \gamma).
\end{equation}
For small $\sigma$ the PBH production rate  $\beta_0$ tends to be proportional to $\sigma^5$, 
\begin{equation} \label{bmat}
\beta_0(\sigma)\, = \, 0.056 \, \sigma^5\,.
\end{equation}  
This expression has been derived with semi-analytical calculations and applies for  $0.005 \lesssim \sigma \lesssim 0.2$, whereas for for $\sigma \lesssim 0.005$  the PBH production rate is modified if there is an angular moment in the collapsing region \cite{Harada:2017fjm}.
For such a small $\sigma$  however, the GW signal becomes negligibly small for the sensitivity of the operating and designed  GW detectors. 
 The PBH dark matter fractional abundance is 
\begin{align} 
f_\text{PBH} \equiv \frac{\Omega_\text{PBH}}{\Omega_\text{DM}}\, \simeq \, 1.3 \times 10^{19} \, \gamma_\text{M}\,  \beta_0\,  \left(\frac{T_\text{rh}}{10^{10}\, \text{GeV}}\right)\,.
\label{ratio}
\end{align}
By definition the fractional abundance 
is equal or less than one but in the largest part of the mass spectrum there are stringent upper bounds on $\Omega_\text{PBH,max}/\Omega_\text{DM}$ due to   observational constraints, see Fig. \ref{PBHplot}. 
The maximum value $\Omega_\text{PBH,max}/\Omega_\text{DM}=1$ is possible in the mass range $M_\text{PBH}\sim 10^{-15}-10^{-10} M_\odot$. 
The $\gamma_\text{M}$ that appears in Eq. (\ref{ratio})  is the fraction of the horizon mass that collapses into PBH during the matter domination era.
It is a numerical factor which depends on the details of gravitational collapse  and it can admit values possibly from $\gamma_\text{M}\sim 10^{-4}$  
up to $\gamma_\text{M}\sim 1$    \cite{Hawke:2002rf, Carr:2009jm}.  Note that PBHs form mostly by spherical collapse and $\gamma_\text{M}$ is not related with the asphericity developed in other Hubble patches that dominate the GWs emission.

According to Eq. (\ref{ratio})  the variance (\ref{bmat})  is bounded from above.
When the upper bound on $\sigma$ is saturated the PBH abundance maximizes.  Several observational experiments determine the  $\Omega_\text{PBH,max}$ 
in a wide range for the mass parameter $M_\text{PBH}$   labeled in Fig.~\ref{PBHplot}. 
Light PBHs are constrained from the extra galactic gamma-ray background (EGB) \cite{Page:1976wx, MacGibbon:1991vc, Carr:1998fw, Barrau:2003nj,  Carr:2016hva}, as well as from $e^{\pm}$ annihilation in the galactic center (GC) and positron constraints (V) \cite{Dasgupta:2019cae}.
Black holes of mass above $10^{17}$g are subject to gravitational lensing constraints  \cite{Barnacka:2012bm, Tisserand:2006zx, Niikura:2017zjd}, given by Subaru (HSC), Ogle (O), EROS (E) and MACHO (M), microlensing of supernova (SN) and others.
The CMB anisotropies measured by Planck (PA) constrains the PBH with mass above $10^{33}$g 
\cite{Ricotti:2007au, Carr:2016drx, Clesse:2016vqa, Bird:2016dcv, Poulin:2017bwe, Serpico:2020ehh}. At the large mass region there are also constraints from accretion limits in X-ray and radio observations \cite{Gaggero:2016dpq} and X-ray binaries (XB) \cite{{Carr:2020gox}}. There are also dynamical limits from disruption of wide binaries (WB) \cite{Quinn:2009zg}, and survival of star clusters in Eridanus II (Er) \cite{Brandt:2016aco}.
Advanced LIGO/Virgo searches for compact binary systems with component masses in the range $0.2-1 M_\odot$ find no viable GW candidates
 \cite{Abbott:2018oah}. 
 The non-observation of the stochastic GW background (GWB) from PBH mergers from LIGO/Virgo constrains PBHs with mass in the range $1-10^2M_\odot$  for the monochromatic mass function
  \cite{Raidal:2017mfl, Hutsi:2020sol}. However disruption of PBH binaries might have been underestimated and it is probable that the merger rate might be smaller thus invalidading the aforementioned constraint Ref.\cite{Jedamzik:2020omx}. 
 Finally second order tensor perturbations (GW2) generated by scalar perturbations already constrain the abundance of PBH masses approximately in the range $10^{-3}- 1 \,M_\odot$. 
For a recent update on the PBH constraints see Ref. \cite{Carr:2020gox}.
What is of special interest in the present work is that the PTA bound\footnote{We mention that the NANOGrav Collaboration has recently published an evidence  of an isotropic stochastic process \cite{Arzoumanian:2020vkk}.} on the second order tensors  (GW2)  \cite{Thorsett:1996dr, Saito:2008jc, Bugaev:2010bb, Chen:2019xse}, depicted in Fig.  \ref{PBHplot}, 
does not apply as it is if a late reheating takes place. Instead, a significant PBH dark matter component can be consistent with the current pulsar timing data
if produced during eMD.

\subsection{PBH as the  entire dark matter of the universe}

In the mass window $M_\text{PBH}= 10^{-15}- 10^{-10}M_\odot$ 
the PBH abundance can reach its maximum value, i.e., it can account for $100\%$ of the dark matter abudnance. 
If PBHs constitute a significant fraction of the total dark matter then a GW counterpart must exist that can be tested by the scheduled Laser Interferomenter Space Antenna (LISA) \cite{Audley:2017drz},
as well as by other space-based  proposed experiments such as  DECIGO \cite{Seto:2001qf, Sato:2017dkf}, BBO \cite{Crowder:2005nr}, 
TianQin \cite{Luo:2015ght}, and Taiji \cite{Guo:2018npi}. 
For (see Eq.~(\ref{reheat}))
\begin{equation}
T_\text{rh}\lesssim 5 \times 10^5  \,\text{GeV} \, \sigma^{3/4}\left(\frac{M}{10^{-12}M_\odot} \right)^{-1/2} 
\label{reheat2}
\end{equation}
our result, given by  Eq. (\ref{result_Om}), applies  and gives the GW spectrum. The GW signal is generally 
found with a different amplitude compared to the signal produced during radiation dominated era and  the peak frequency  shifted towards smaller values as the reheating temperature decreases (see Fig. \ref{Fig:TempLISA}).
In view of the $\gamma_\text{M}$ uncertainties, we  plot in Fig. \ref{Fig:TempLISA}
the GW signal for two different $\gamma_\text{M}$ parameters in the LISA-BBO frequency band. For the  PBH dark matter scenario, smaller $\gamma_\text{M}$ parameter values increase the GWs amplitude.  The peak frequency of the signal is the signature of the PBH dark matter scenario in cosmologies with low reheating temperatures.

\subsection{PBH with solar and sub-solar masses}

Besides the recent result  of NANOGrav Collaboration  \cite{Arzoumanian:2020vkk}, 
PTA experiments rule out the existence of practically any PBH with mass $M_\text{PBH}={\cal O}(10^{-3}-1) M_\odot$ \cite{Thorsett:1996dr, Saito:2008jc, Bugaev:2010bb, Chen:2019xse}, as depicted in Fig.  \ref{PBHplot}. 
However, the depicted PTA bound  on the second order tensors (GW2)  has been derived assuming a radiation dominated era.
For reheating temperature  (see Eq.~(\ref{reheat2})) less than $T_\text{rh} <   10^3  \sigma^{3/4}(M_\odot/M)^{1/2} \,\text{MeV}$,
 GWs  that are produced during eMD might avoid the severe  GW2 constraint on the PBH abundance, depicted in Fig.  \ref{PBHplot}.
Hence, at that mass range $10^{-3}- 1 \,M_\odot$  the PBH abundance can be a non negligible fraction of the total dark matter abundance, even up to few $\%$ according to the bounds from EROS and MACHO microlensing experiments, 
without overproducing  GWs.
 Let us note that a further motivation in this mass range is the intriguing event \cite{Abbott:2020khf} and the recent evidence for a  stochastic GW background \cite{Arzoumanian:2020vkk}.

\begin{figure}[!htbp]
  \begin{subfigure}{.5\textwidth}
  \centering
  \includegraphics[width=.95 \linewidth]{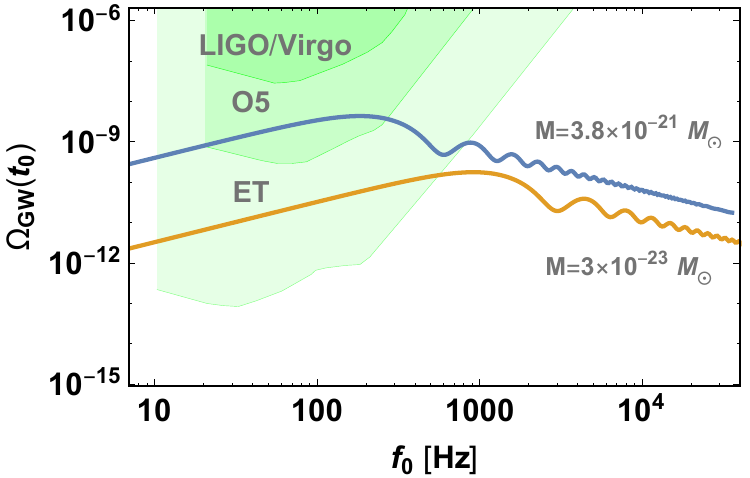}
\end{subfigure}  
\begin{subfigure}{.5\textwidth}
  \centering
  \includegraphics[width=.95\linewidth]{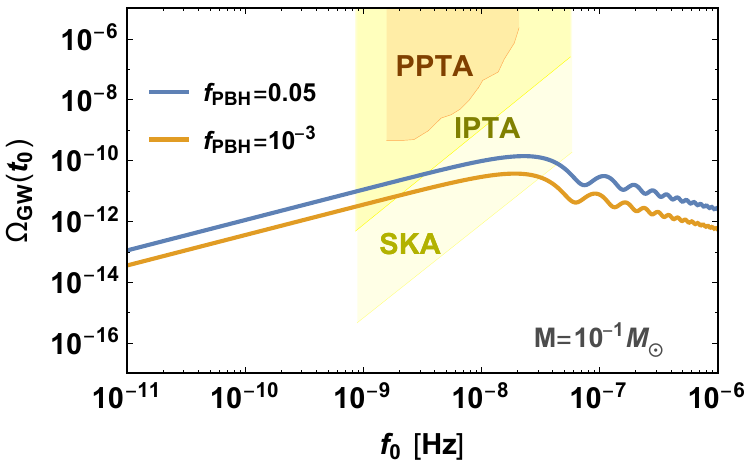}
\end{subfigure} 
 \caption{\label{LigoRem}~~  
{\it Left  panel.} The GW signal in the LIGO/Virgo and Einstein Telescope frequency range for promptly evaporating PBHs and $\sigma=0.1$, $T_\text{rh}=2\times 10^9$ GeV
for two horizon masses, $ M=3.8\times 10^{-21} M_\odot$ and  $ M=3\times 10^{-23} M_\odot$. The BBN constraints are evaded for $10^{-4} \lesssim \gamma_\text{M}\lesssim 2\times10^{-2}$. If these PBHs leave behind a remnant with mass $\kappa M_\text{Pl}$ it can constitute a significant part of the dark matter in the universe for $\kappa\gg 1$.
{\it Right Panel.} The GW signal in the PTA frequency range, for horizon mass $M=10^{-1} M_\odot$ and for two fractional abundances for the associated PBHs   $5\%$  and  $0.1\%$ and for $\gamma_\text{M}=1$. 
 The $\sigma$ value corresponds to $0.023$ (upper curve) and $0.012$  (lower curve) and the reheating temperature is chosen to be $T_\text{rh}=T_\text{col}/2$, that is $87$ and $53$ MeV respectively.
  }
\end{figure}

\begin{figure}
     \centering
     \includegraphics[width=0.8\columnwidth]{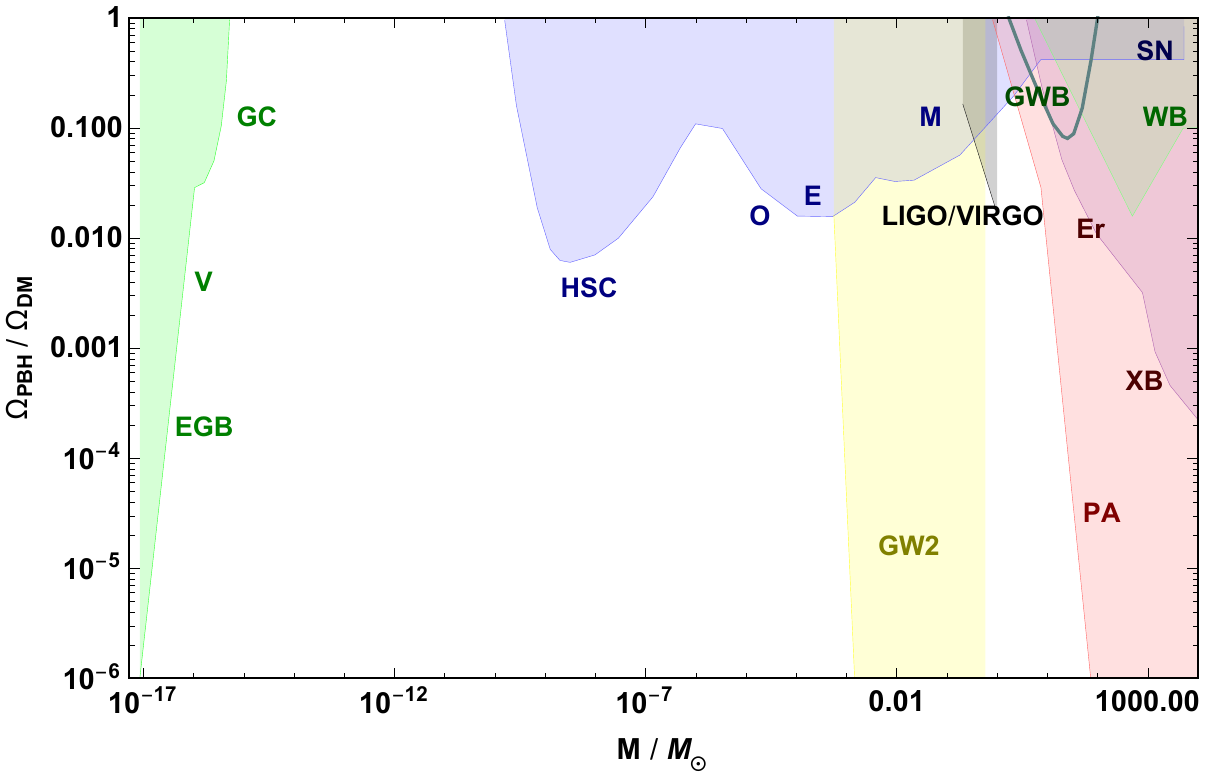}
     \caption{The observational bounds on $\Omega_\text{PBH}/\Omega_\text{DM}$ \cite{Carr:2020gox}, as described in the text. Note that the depicted GW2 bound does not apply in our case.}
     \label{PBHplot}
 \end{figure}

As an example, let us take $\Omega_\text{PBH}/\Omega_\text{DM}=5\%$ for $M_\text{PBH}=0.1 M_\odot$ and $\gamma_\text{M}=1$.  
According to   Eqs.~(\ref{bmat}),  and (\ref{ratio}) such an abundance is obtained for $\sigma=0.023$. 
For $T_\text{rh}=1/2T_\text{col}=87$ MeV, these parameters yield a GW signal that has a maximal amplitude  $\Omega_\text{GW}(t_0, f_\text{peak})=1.4\times10^{-10}$ at the peak frequency $f_\text{peak}=2.4\times 10^{-8}$ Hz, see Eq. (\ref{Omegapeak}) and (\ref{fMDpeak}), and this value is fully compatible with the current PTA constraints. In Fig. \ref{LigoRem} the
 GW signal has been plotted together with a second case that yields $\Omega_\text{PBH}/\Omega_\text{DM}=0.1\%$.
Near future PTA experiments, SKA \cite{Janssen:2014dka, Maartens:2015mra} and FAST \cite{Lu:2019gsr}, can test the PBHs scenarios in eMD cosmologies with low reheating temperatures and with  mass in the range between the earth and the sun. 

\subsection{Ultra light PBHs }

Lighter PBHs are associated with GW production in larger frequencies. 
Currently the LIGO-Virgo probes the frequency band around  $f=100$ Hz. 
According to Eq.~(\ref{fMDpeak}) GWs will be produced  with that frequency range  if PBHs have mass $\gamma_\text{M} M_\text{PBH}\sim  10^{-20} M_\odot\, (T_\text{rh}/10^9 \text{GeV})$. 
LIGO-Virgo, as well as the Einstein Telescope \cite{Sathyaprakash:2012jk},  can hence detect GWs  associated with PBHs  with mass $10^{-20} M_\odot$ or less, see e.g. \cite{Dalianis:2020cla, Bhaumik:2020dor}.
PBHs with very light masses are anticipated to Hawking radiate and evaporate in timescales less than the age of the universe for $M_\text{PBH}<10^{-16} M_\odot$. Due to the energetic Hawking radiation, strong constraints on their abundance have been placed \cite{Carr:2009jm, Carr:2020gox}.
However, the scenario of mini PBHs is interesting due to the theoretical expectation that black holes might not evaporate into nothing but leave behind a stable state, called black hole remnant.
PBH remnants can actually comprise the entire dark matter of the universe \cite{Barrow:1992hq, Carr:1994ar, Dalianis:2019asr}. For PBH formation during a matter domination era the fractional abundance of the PBH remnants is \cite{Dalianis:2019asr},
\begin{align}
\frac{\Omega_\text{rem}}{\Omega_\text{DM}}
\simeq  3 \times  \,10^{-6}\, \kappa  \left(\frac{\beta_0}{10^{-9}}\right)  \left( \frac{M}{10^{10} \text{g}} \right)^{-1 }
\left(\frac{T_\text{rh}}{10^{10}\text{GeV}} \right),
\end{align}
where $M_\text{rem}=\kappa M_\text{Pl}$ is the mass of the remnant. 
PBHs with mass less than about $10^9$ grams are free from the BBN observational constraints  \cite{Carr:2020gox}. 
The maximum value of the GW signal is $\Omega_\text{GW} \sim 4.5\times 10^{-22} \sigma (T_\text{rh}/\text{GeV})^2(f_\text{peak}/\text{Hz})^{-2}$.  Assuming small values for the $\gamma_\text{M}$ parameter the GW signal can be strong enough to be 
detectable by LIGO-Virgo without violating  BBN constraints.
Particular combinations of $\kappa$ and $\gamma_\text{M}$  values can also realize the PBH remnant dark matter scenario
and near future experiments could set constraints on the anticipated GW signal, see Fig. \ref{LigoRem}.

\section{Conclusions} 
\label{secConclusions}

 In this work we examined the gravitational waves  produced during an early matter domination era  due to the non-spherical evolution of the primordial density perturbations. 
We applied the Zel'dovich approximation to describe  the nonlinear evolution of the density perturbation  and derived the expression for the GW spectrum by exerting the quadrupole piece. 
 To our knowledge, this is a novel result since this is the first time where this non-perturbative method is used in order to estimate the GW signal. 
After a series of analytical steps 
 we estimate the GW signal numerically, performing the full integration of the triple integral in Eq. (\ref{result_Om1}).
 This paper also serves as a proof of concept.  Within our method it is allowed to make a full very precise non-perturbative estimate 
increasing $N$ the number of time intervals which we partition the time interval $[t_\text{max},t_\text{col}]$.

We have found  distinct  features for the GW signal which can distinguish it from GW signals of a radiation domination scenario: it has a different amplitude, peaked at   different frequencies shifted towards smaller values as the reheating temperature decreases  and with different power-law scaling compared to the radiation domination case.  
Turning to the PBH counterpart we also find different predictions.  PBHs can be the dark matter of the universe  in the mass region $M_\text{PBH}\sim 10^{-15}-10^{-11}M_\odot$ with the associated  GWs signal being inside the sensitivity curve of the LISA curve. 
In addition, PBHs in the solar mass range or less, $M_\text{PBH}\sim 10^{-3}-1M_\odot$, can have a significant abundance without violating  the existing PTA constraints (where models of PBH  formed in radiation domination  are subjected to), if the reheating temperature is sufficient low.

We plan to extend our work also by studying the GW signal produced from overdensities that did not form black holes as they evolve from a collapsed structure to a virialized halo. This signal could be as strong as the one estimated here and it could have different spectral characteristics which could facilitate the identification of a potential signal easier.

It is expected that gravitational-wave astronomy will make a significant progress in the next decades. This could allow us to probe and test non-thermal early universe histories as well as PBH dark matter scenarios. In addition, the results  of the present work contribute to the understanding of  the properties and  variations of the presumably present stochastic gravitational wave background.

\section*{Acknowledgments}
I.D. would like to thank CP3 Origins, University of Southern Denmark  for hospitality during the early stages of this work.
The research work of I.D. was supported by
the Hellenic Foundation for Research and Innovation (H.F.R.I.) under the
"First Call for H.F.R.I. Research Projects to support Faculty members and
Researchers and the procurement of high-cost research equipment grant"
(Project Number: 824).

\end{document}